# Direct Thermal Imaging of Domain Wall Hot Spots in $LiNbO_3$

*Lindsey R. Lynch*, J. Marty Gregg, Amit Kumar, Kristina M. Holsgrove and Raymond G. P. McQuaid**

Centre for Quantum Materials and Technologies, School of Mathematics and Physics, Queen's University Belfast, University Road, Belfast BT7 1NN, United Kingdom
E-mail: llynch14@qub.ac.uk, r.mcquaid@qub.ac.uk

Funding: UKRI Future Leaders Fellowship (grant no. MR/T043172/1), US-Ireland R&D Partnership Programme (grant no. USI-205)

Ferroelectric domain wall devices offer a promising route to low voltage, reconfigurable nanoelectronics by confining currents to nanoscale conducting interfaces within an insulating bulk. However, the potential for resistive heating and unregulated temperature increases due to domain wall conduction remains unexplored. Here, we employ scanning thermal microscopy to directly image hot spots in thin-film lithium niobate domain wall devices. Piezoresponse force microscopy shows that the hot spots correlate with nanodomain structure and thermal mapping reveals surface temperature rises of $\sim 20$ K at most, levels that are unlikely to negatively affect device performance. This is due to the moderate electrical conductivity of domain walls, their voltage-tunable erasure, and distributed current pathways, which inherently limit power dissipation and peak temperatures. Finite element electrothermal modelling indicates that domain walls behave as pseudo-planar heat sources, distinct from the filament-based heating typically observed in resistive switching oxides. These findings highlight the potential for domain wall devices as an energy-efficient, thermally stable platform for emerging memory and logic applications.

## 1. Introduction

For over a decade, electrically conducting ferroelectric domain walls (DWs) have drawn attention as reconfigurable functional elements that could be used in novel storage memories and neuromorphic device applications. DWs are interfaces that develop naturally at intersections between polar domain variants and their position can be manipulated through well understood voltage-controlled domain nucleation and growth processes.[1,2] DWs are particularly interesting because they can exhibit greater electrical conductivity than the surrounding bulk, facilitated through extrinsic defect engineering of the wall itself, or by introducing a polarization discontinuity that requires compensation by mobile charges.[3,4] Although the electrical transport characteristics of conducting walls often appear thermally activated, this remains under some scrutiny,[5] with metallic-like,[6] and even superconducting, behaviors having been reported.[7] When integrated into demonstrator devices, such as transistors and memristors,[8–11] operation involves using DWs to reversibly make/break contact with electrodes, performing an analogous role to the vacancy filaments that enable resistive switching in transition metal oxides, such as e. g. $TiO_2$,[12] $HfO_2$,[13] and $TaO_x$[14].

Of the ferroelectric systems that are currently known to exhibit enhanced DW electrical conductivity, lithium niobate shows the largest conductance contrast between DWs and bulk.[15] This is primarily because the bulk is highly electrically insulating, while the walls have best-estimate conductivity values in the range of intrinsic semiconductors.[15–17] Nonetheless, $LiNbO_3$ serves as an ideal platform for testing DW device prototypes because it effectively confines device currents to the walls, allowing functionality to be derived entirely from the number density of walls connecting the electrodes. This is not typical, since other ferroelectric systems displaying enhanced DW conductivity can also have non-negligible bulk conductance, meaning that electrical current pathways are not necessarily well controlled.[18] As a result, DW enabled rectification,[15,17] transistors,[19,20] memristors,[21,22] and logic gates,[20,23] have been most convincingly demonstrated using $LiNbO_3$ integrated into various device geometries. Although most studies have focused on electrical characterization, power dissipation in the current carrying DWs will also cause localized Joule heating, which could impair performance or lead to thermal crosstalk in high-density memory arrays. This is already a known issue for resistive switching in transition metal oxides, where vacancy filament formation and conduction are fundamentally electrothermal processes which depend on self-heating and local temperature conditions.[24,25] However, studying the fundamental properties of the filaments directly is challenging since they can be truly nanoscale, with diameters less than 10 nm. Several factors complicate the direct measurement of filament temperatures via surface thermometry, including lateral heat spreading in the active layer and electrodes, as well as increased sensitivity to interfacial thermal resistances in nanoscale devices.[26,27]

Over the last few years, the nanoscale spatial resolution of scanning thermal microscopy (SThM) has made it an invaluable tool for direct thermal imaging of filamentary behavior in resistive switching devices. Deshmukh *et al*.[13] were first to investigate local Joule heating of a single filament by SThM in $HfO_2$ resistive random access memory devices (RRAMs), revealing surface temperature 'hot spots' of over 350 °C, and even higher implied filament temperatures of ~1300 °C, due to nanoscale confinement of dissipated power. Similar local heating investigations by SThM have subsequently been carried out for $NbO_x$ memristors,[28] $TiO_2$ RRAMs,[12,29] and on resistive switching devices based on 2D $MoTe_2$.[30] As well as revealing the location of filamentary current pathways indirectly through local heating, the studies explore how filament heat generation and spreading can affect performance in realistic device geometries. In this study, we use a combination of SThM and piezoresponse force microscopy (PFM) domain imaging to reveal and quantify the resistive heating associated with ferroelectric DWs under steady-state current in thin film $LiNbO_3$ devices. Using SThM, we directly observe hot spot features with surface temperature rises ranging from sub-Kelvin up to ~20 K, depending on the dissipated power levels. Unlike in conventional resistive switching systems, we are able to leverage PFM to directly image the distribution of heat sources and correlate this with the observed thermal signals to better understand the relationship between microstructure and device heating.

## 2. Results and Discussion

### 2.1 Device Fabrication and Domain Wall Engineering

DW memristor devices (schematized in **Figure 1**a) were fabricated using commercially available z-cut thin films of $LiNbO_3$ that have been ion-sliced from congruently grown single crystals and then thermally bonded onto prepared carrier substrates. The structure consisted of 500 nm $LiNbO_3$/150 nm Cr-Au-Cr/2 μm $SiO_2$/0.5 mm $LiNbO_3$. In order to engineer the desired microstructure, electrically conducting DWs were injected into the as-received monodomain $LiNbO_3$ thin film layer by an established poling process.[21,22] This involved local contacting and biasing of the bare film surface with a removable In-Ga-As liquid top electrode. Since the $LiNbO_3$ bulk material is highly-insulating, it is reasonable to assume that any steady-state current is due to charge transport along conducting DWs that connect the device electrodes. Therefore, we relied on current feedback to determine the applied bias required to maximize the number of DWs injected under the liquid electrode. An example of the typical domain microstructure obtained by this process is revealed in PFM and conductive-atomic force microscopy (c-AFM) mapping of the bare surface, shown in Figure 1b. Regions of densely packed nanodomains are revealed and spatially resolved current mapping (Figure 1b) shows enhanced electrical conductivity within the polydomain region. This is consistent with previous reports in lithium niobate,[21,22,31,32] where enhanced conductivity is attributed to mobile charge transport along charged DWs. More broadly, enhanced electrical conductivity along DWs in ferroelectrics is a well-established phenomenon which can be explained by a combination of defect-aggregation enabled transport, band-gap reduction at walls, or by compensatory charge accumulation at electrostatically charged interfaces.[4] The latter mechanism is key for z-cut lithium niobate thin films, where intermediate switched states involve high spatial densities of charged needle-like domains, which have previously been described as being like truncated cones due to the large angle of inclination (~10°) of the DWs.[16] This domain morphology is consistent with that expected for the transition from the forward growth phase[2] of switching to the early stages of sideways growth. Importantly, since our c-AFM mapping indicates that current can be passed through DWs, this means that the tips of the domain needles have reached the opposite electrode and formed continuous conducting pathways.[22] It is important to note that it is solely the DWs (interfaces) that carry current, and not the domain volume. The observed smearing of current in Figure 1b is due to limitations in imaging resolution and further current mapping is shown in Figure S1 to emphasize that the signals originate from the walls. For the device, a long platinum top electrode of 35 nm thickness and lateral dimensions of 62.5 μm by 1 mm was deposited on top of the prepared domain structure for device measurements and thermal investigation. The domains spanned a 160 μm long region under the electrode, demarcated by the white lines in **Figure 2**a. Therefore, when the top electrode is biased at one end, charge flow is confined to within the bar (since the monodomain film contacted underneath is highly insulating) until the DWs are reached, which provide leakage paths to the grounded bottom electrode (see Figure 1a inset). Since the DWs in thin-film $LiNbO_3$ are considered to be semiconducting,[16,17] they should present the highest electrical resistance contribution in the circuit. Therefore, there is an expectation for localized Joule heating within the DWs that pass current between the top and bottom electrodes. Forward-bias current-voltage measurements of the prepared DW device are shown in Figure 1c, with steady-state currents measured up to 100 μA. These currents are maintained at a low level to maintain device integrity and to avoid electric field induced domain switching.

## 2.2 Initial observations of local heating in $LiNbO_3$ Domain Wall Device

To obtain an initial overview of the device self-heating, SThM was carried out over the length of the top electrode while under a current of 60 μA (0.66 mW power), with the topography map of the region shown in Figure 2a. The raw temperature data was background corrected (see Figure S2 and Note 1) to minimize sporadic noise that arises from changing tip-surface conditions during the scan, resulting in the temperature profile shown in Figure 2b. A region of elevated temperature is clearly observed, which lies within the central area that was prepared with domain microstructure (demarcated by the white dashed lines). This would strongly suggest that current leakage through the DWs, and associated Joule heating, is responsible for the detected hotspots. The decay in temperature away from the center to either end of the electrode further implies that the buried DWs are the main source of heat in the device. Taking a line section of surface temperature along the center of the electrode long axis (Figure 2c) details a primary "hot spot" feature with a modest local relative temperature rise of ~0.3 K. A neighboring hot spot can also be identified, which is smaller in area and has a smaller temperature rise of less than 0.1 K. The peak temperature of the larger hot spot scales linearly with the dissipated power (determined using two-probe current and voltage readings), adding confidence that the observed signals have their origin in Joule heating (see Figure 2d). To identify if self-heating of the top electrode is a significant contributor to the observed temperature increases, we also measured Joule heating for currents directed solely through the top electrode (see Figure S3), finding this to be a negligible contribution compared to resistive heating by the DWs. We have also observed larger hot spot peak temperatures of just over 20 K for proportionately larger dissipated power (~ 35 mW) in another similarly prepared DW device, shown in Figure 2e,f and further discussed in Figure S4.

## 2.3. Correlation between Domain Structure and Hot Spot Formation

To investigate the relationship between the spatial distribution of DWs and observed heating signals, we carried out polar domain mapping surrounding the hot spot location for the device shown in Figure 2a,b. Higher resolution thermal mapping of the hot spots is shown in **Figure 3**a, with through-electrode PFM of the same region in Figure 3b. We extracted contours of the perimeter of written domains from the PFM, and superimposed these on the temperature map, showing that the temperature peaks previously identified in Figures 2b and c can be associated with underlying domain microstructure in the ferroelectric thin film. Higher resolution imaging of the polar microstructure in these regions (Figure 3c) reveals a densely packed arrangement of nanodomains, which were not immediately apparent from the lower resolution survey scan in Figure 3b because of their small size, with diameters of 100 - 150 nm. The nanodomain morphology suggests that the walls have a through-depth conical aspect, as indicated by cross-sectional electron microscopy in previous investigations of conducting DWs in these commercial thin films.[16,22]

In **Figure 4**, we examine the spatial development of the main DW hot spot with increasing device power. The temperature map for the device in the power off state is shown in Figure 4a and local heating is mapped for increasing dissipated power on the order of 0.1 mW in Figure 4b-e. Initially, as the bias is increased above 0 V, local heating begins to occur and the hot spot becomes only faintly observable above background (by ~60 mK in Figure 4b), likely due to the prevalence of heat spreading into the surrounding material. It is worth noting that the heating observed at these lower voltages still originates from power dissipated in the DWs and not due to self-heating of the top electrode (see Figure S3). As power is increased further (Figure 4d and e), the localized heating becomes more obvious and the temperature maps begin to display hot spot features that resemble the perimeter of domains seen in the PFM map (overlaid in figure 4e). This behavior is also captured in the Figure 4g temperature line sections, which show the hot spot emerging above background during a slow D.C. voltage sweep, where enough time was given for the tip-sample system to thermally equilibrate at each step. Comparing the PFM and SThM maps, we see that there is not a one-to-one match between the domain microstructure and the heating signals, even at the highest power. A lack of hot spots in regions where domain coverage is sparse may be suggestive of a critical DW density and threshold for local power dissipation to generate detectable hot spots. Alternatively, these DWs may not actually be carrying electrical current, due to poorly formed domain structures that do not connect the electrodes or inherent complexity in the current percolation pathway.[33] Changes in the hot spot morphology with increasing voltage are primarily due to increases in the locally dissipated power, not by bias-induced changes in the microstructure (i.e. nucleation and growth of polar domains). We verify this by PFM domain imaging carried out before and after the thermal mapping, which shows no noticeable changes in the microstructure (Figure S5). Nonetheless, it is important to note that polarization switching can be triggered by sufficiently large voltages (> 25 V) and will place an upper limit on the achievable power density and associated self-heating that can be achieved. This is facilitated by domain growth and coalescence, which reduces the DW number density, and results in a progressively increasing device resistance until switching is complete and no current leakage pathways remain.

**2.4. Electrothermal modelling and implications for device performance**

To rationalize the steady-state temperature patterns observed experimentally by SThM, we have carried out finite element electrothermal modelling where the PFM domain maps were used as input to recreate the DW distribution (for details, see Experimental Methods). We modelled the DWs as 2D current-carrying interfaces and used DW electrical conductivity as a fitting parameter to match the experimentally measured device resistance. The modelled surface temperature map for the biasing conditions in Figure 4e is shown in Figure 4f and further finite element modelling is detailed in Figure S6. Even though the heat originates from a nanostructured arrangement of DWs, the resulting modelled hotspot temperature distribution varies smoothly on the microscale and the main features are captured in our SThM maps. The good qualitative match between model and experiment therefore indicates that the DWs can reasonably be considered to behave as planar heat sources buried within the surrounding inactive bulk, which is conceptually distinct from filament-based models used to describe resistive switching in binary oxides. For the device geometry studied, there is no significant difference between the interior temperature of the DWs and the top electrode surface temperature even when reasonable values for interfacial thermal resistances between the electrodes and $LiNbO_3$ are assumed (see Figure S7 and Note 2). To explore the effect of downscaling on the running temperature, a 1x1 μm device is modelled in Figure S8, which results in an average current density of ~118 nA/$\mu m^2$ for 5 V operation (comparable in performance to $Pb(Zr,Ti)O_3$ DW devices).[34] In this case, the dissipated power of 0.59 μW is associated with only a small peak temperature rise of ~20 mK. From a practical point of view, it therefore seems unlikely that the large temperature increases seen in filamentary systems (potentially up to $10^3$ K)[13,25] would arise in DW memristors, unless devices can be configured for high-power operation through further enhancements in DW electrical conductivity (Figure S6d), or optimized device designs.[35] This is necessary because deletion of DWs at super-coercive biases typically places an effective upper limit on the current that can be forced through the device. In addition, the web-like geometry of DWs helps to distribute current and avoid thermal bottlenecks, as can occur in isolated filaments due to extreme power confinement. The expectation for low running temperatures has positive implications for reliability, especially since $LiNbO_3$ DW currents can become unstable and decay in time if temperature is elevated above 70 °C.[15] From a power efficiency standpoint, the use of $LiNbO_3$ is advantageous because device electrical conductance can be tuned by minor adjustment of wall inclination angle using sub-coercive reverse bias, therefore enabling reproducible resistance control without resorting to energy-inefficient polarization cycling.[21]

## 3. Conclusion

In conclusion, we have used SThM to spatially map the local heating that occurs in $LiNbO_3$ DW memristor devices under steady-state current. We find clear evidence of local hot spots developing on the electrode surface, with peak temperature rises ranging from sub-Kelvin to 20 K, proportionate to the magnitude of device current. By using PFM to image the polar microstructure directly, we verify that these heating signals are due to power dissipated within the conducting DWs. However, not all observed nanodomain regions are associated with hot spots, due to either insufficient local heating or because some DWs do not contribute to electrical current leakage. Electrothermal finite element modelling shows that the mapped surface temperature distribution can be broadly reproduced by treating the DWs as pseudo-planar heat sources, which is qualitatively distinct from that of filamentary systems. DW devices offer unique advantages through structural tunability and potential for low-voltage operation, while their comparatively low thermal footprint minimizes the risk of thermal degradation and crosstalk. Furthermore, combined SThM and PFM investigations (also demonstrated in[36]) can provide direct thermal-structural correlations that are not easily accessed in conventional filament-based systems. Finally, deploying DWs as configurable planar heaters may open up opportunities for customizable microscale heating,[37] with potential applications in nanomaterials synthesis,[38] particle manipulation,[39,40] and electrothermal actuation.[41] In such contexts, spatially selective heating via voltage-controlled DW patterning could offer a lithography-free alternative for fabricating complex heater arrays.[42]

## 4. Experimental Methods

*Piezoresponse Force Microscopy and current-AFM mapping*: All scanning probe microscopy measurements were carried out using an Oxford Instruments MFP-3D Asylum Atomic Force Microscope. For PFM and c-AFM, Pt/Ir-coated Si probes (Nanosensors model PPP-EFM tips) were used, with 2 $V_{ac}$ potential difference applied to the probes at 330 - 340 kHz for PFM and 10 $V_{dc}$ for c-AFM.

*Scanning Thermal Microscopy:* Spatially resolved temperature mapping of the self-heated device was carried out using SThM mode on an Oxford Instruments MFP-3D Asylum Atomic

Force Microscope. Kelvin Nanotechnology model KNT-SThM-2an probes were used, which comprise of a silicon nitride cantilever with a palladium track micropatterned onto the tip. As the tip is rastered over the surface, maps of tip resistance are obtained using an in-built Wheatstone bridge sensing circuit, which can be converted into maps of surface temperature using a calibration factor determined from SThM measurements made on an independently temperature-controlled Pt surface. Further details on the tip calibration are presented in Figure S9 and Note 4.

*Electrothermal finite element modelling:* Finite element modelling was carried out for a 3D geometry using the Electric Currents and Heat Transfer interfaces in COMSOL Multiphysics software package. The modelled thin film layer thicknesses matched those of the sample used in the experiment and the Pt bar dimensions were chosen as 62.5 x 193 x 0.04 $\mu m^3$. The modelled substrate dimensions of 260 x 140 x 50 $\mu m^3$ were chosen as a tradeoff between recreating the real bulk geometry and considerations of computational expediency. All external geometry faces were assigned adiabatic heat flow conditions, except for the substrate base which had a fixed temperature condition. COMSOL material files were used as input for the electrical and thermal properties of $LiNbO_3$, $SiO_2$ and the buried Cr-Au-Cr electrode was treated as elemental Au. The electrical and thermal conductivities of the Pt thin-film top electrode were reduced from elemental values by a factor 10, based on two-probe resistance measurements and assuming adherence to the Weidemann-Franz law. The DWs were modelled as internal interfaces assigned with the 'electrical shielding' boundary condition, enabling 2D electrical currents between the top and bottom biased electrodes. The electrical conductivity of the DW interfaces was used as a fitting parameter so that the modelled current and voltage values (and hence power) matched the experimentally measured ones. This was achieved by assuming a DW conductivity of $\sigma_0 = 0.35$ S/m. This value for $\sigma_0$ should not be overinterpreted, as it depends on assumptions about DW density and subsumes the effect of any electrical contact resistances between the electrode layers and $LiNbO_3$ thin film. The COMSOL heat transfer interface was used for generating 3D temperature maps associated with the Joule heating, from which surface temperature profiles on the Pt electrode and within the $LiNbO_3$ layer were extracted. The recreation of the hot spot morphology observed by SThM is encouraging (Figure S6), considering the range of assumptions made in the model, including: the thermal properties of each layer, the spatial density and morphology of DWs, the DW electrical conductivity, no electrical interfacial resistances accounted for, or any possible confounding role played by the SThM tip.

**Supporting Information**

Supporting Information is available from the Wiley Online Library or from the author.

**Acknowledgements**

This work was supported by a UKRI Future Leaders Fellowship (grant no. MR/T043172/1). The Department for the Economy (NI) is acknowledged by R.G.P.M. and L.R.L. for postgraduate studentship funding and by K.M.H. and A.K. for support through the US-Ireland R&D Partnership Programme (grant no. USI-205). L.R.L. and R.G.P.M. acknowledge support by the Lorna Clements Studentship. R.G.P.M. and L.R.L. acknowledge experimental guidance on $LiNbO_3$ and SThM by Dr Conor McCluskey and Dr Rebecca Kelly at QUB. R.G.P.M. and L.R.L. would like to thank Dr Nele Harnack and Dr Bernd Gotsmann (IBM Zurich) for useful discussions on data interpretation and SThM.

**Conflict of Interest**

The authors declare no conflict of interest.

**Author Contributions**

L.R.L. performed all of the sample preparation, scanning probe microscopy experiments and finite element modelling. The study was conceived of by R.G.P.M.. R.G.P.M. and A.K. realized the SThM capability needed for the experiments. R.G.P.M. and K.M.H. supervised the research. The manuscript was written mainly by R.G.P.M. and L.R.L., who were also primarily responsible for data analysis and interpretation. All authors contributed to the discussion and interpretation of results, and were involved in the manuscript editing.

**Data Availability Statement**

The data that support the findings of this study are openly available at the following URL/DOI: https://doi.org/10.17034/e314054a-1735-405f-9db6-d1a9b2131150

## Figures

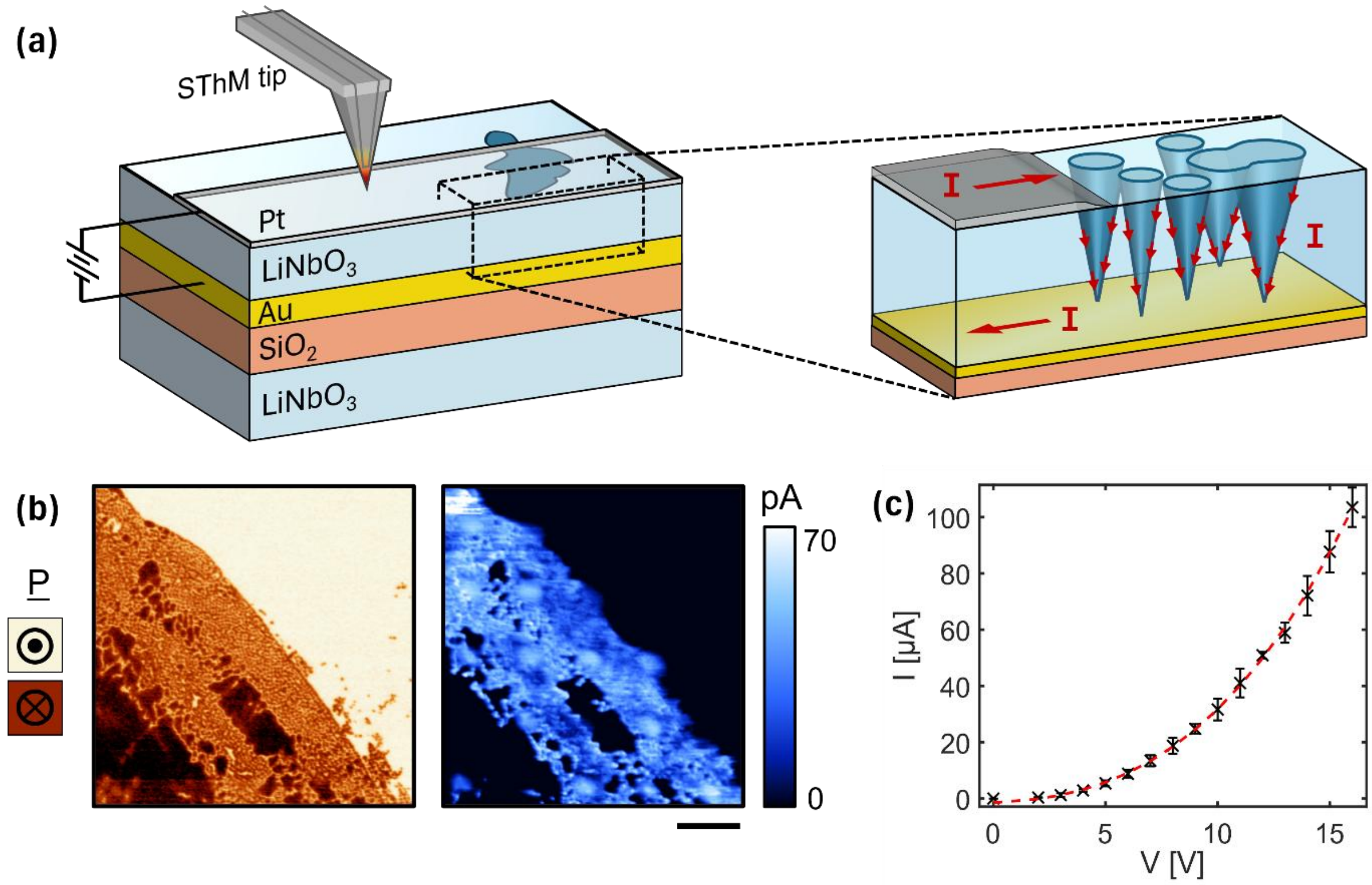

**Figure 1.** (a) Schematic of $LiNbO_3$ domain wall device under bias. Scanning thermal microscopy is carried out over the Pt top electrode. Zoom-in schematically shows the conical domains formed in the top thin-film $LiNbO_3$ layer, with domain walls acting as the only conducting pathways connecting top (Pt) and bottom (Cr-Au-Cr) electrodes. (b) Typical maps of piezoresponse phase (left) and current (right) that are obtained for polydomain regions on the bare $LiNbO_3$ surface. Enhanced conduction is only observed within polydomain regions. Scale bar represents 3 µm. (c) Measured I-V characteristic across the top/bottom electrodes of the domain wall device.

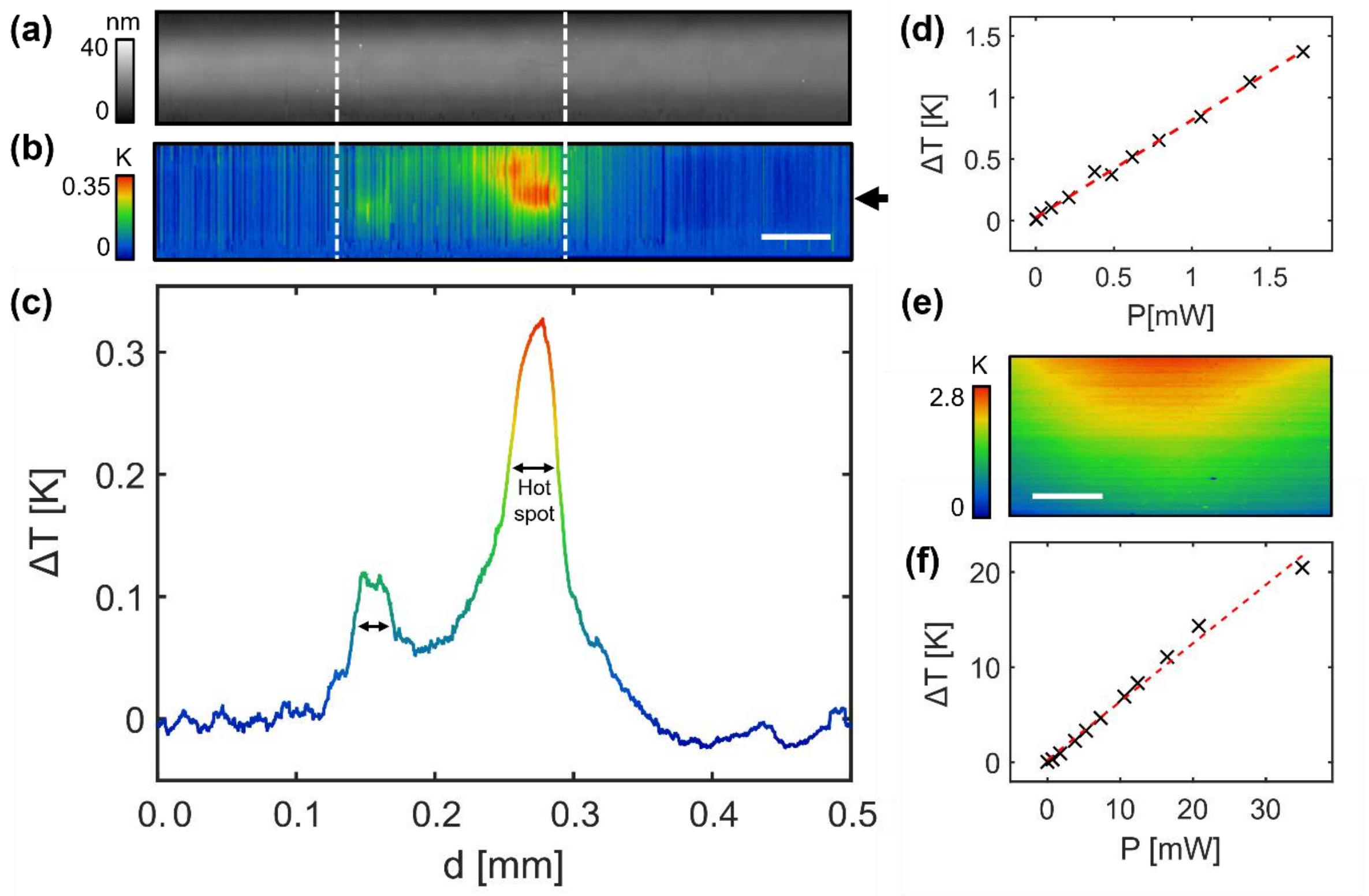

**Figure 2.** (a) 2D topographical map of the Pt electrode deposited on the top surface of the $LiNbO_3$ film. The white dashed line demarcates the approximate region on the $LiNbO_3$ thin film that was prepared with domain microstructure. (b) 2D temperature map obtained using scanning thermal microscopy with device under a constant current of 60 µA. The temperature map was background corrected to remove line noise, highlighting hot spots above the regions of the $LiNbO_3$ sample where domain walls were injected. The scale bar measures 50 µm. (c) A 1D temperature line profile taken along the long axis of the electrode, indicated by the arrow in (b). The line profile was taken from the background corrected map, (b), and a 4-point moving average was then applied. The two hotspots are clearly revealed and are indicated by double headed arrows. (d) Peak absolute temperature rise of the main hot spot in (b)-(c), as a function of power dissipated through the device. (e) SThM map of a hot spot observed in a different device under a dissipated power of just over 3 mW. The scale bar represents 20 µm. (f) Peak absolute temperature versus power for the hot spot in (e).

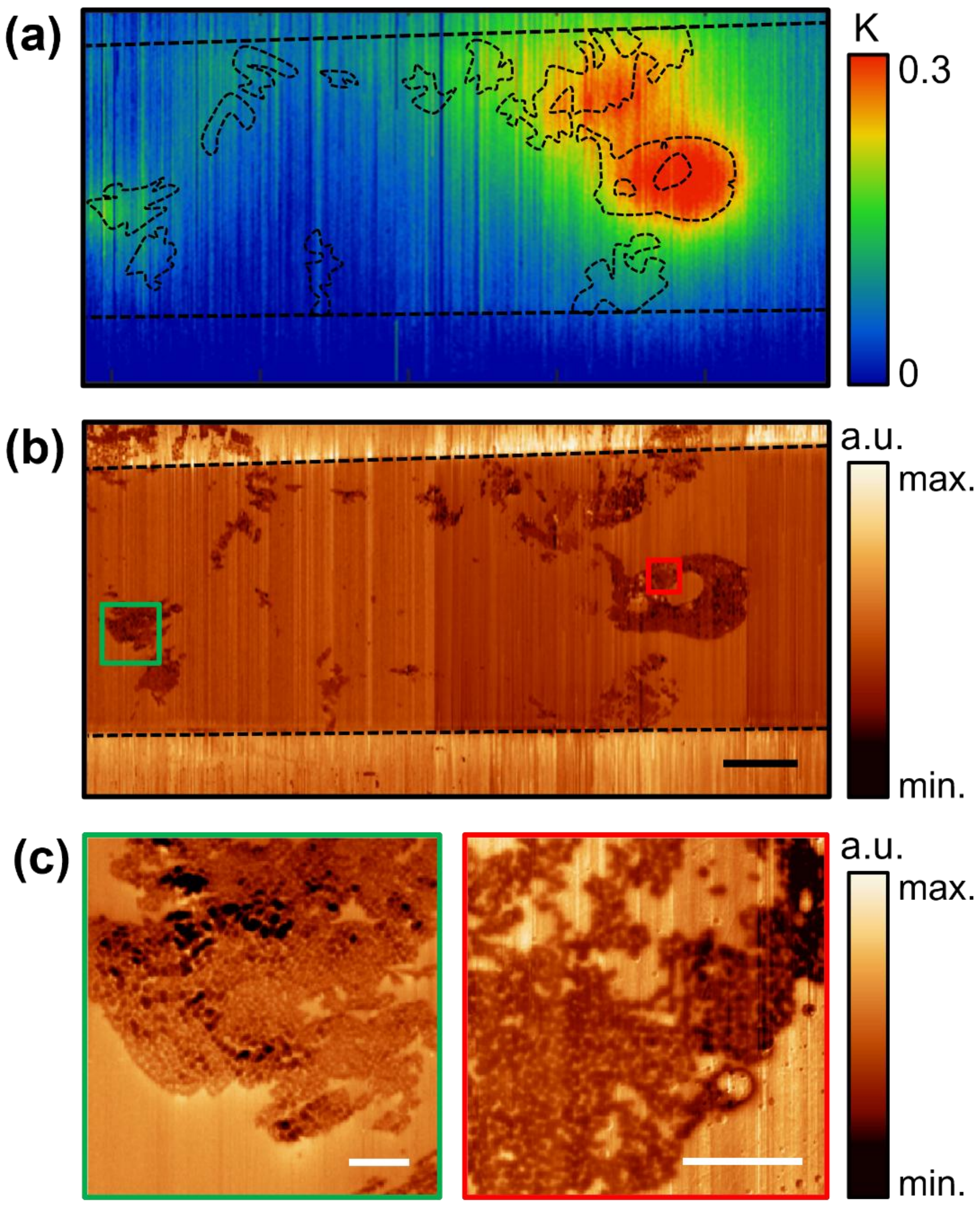

**Figure 3.** (a) The hotspot region mapped using scanning thermal microscopy, overlaid with the perimeter of the partially switched domain regions extracted from through-electrode piezoresponse force microscopy data shown in (b). The horizontally running dashed black lines indicate the edges of the top electrode. The locations of the hotspots correspond to polydomain regions in the $LiNbO_3$, indicating that the elevated temperatures are due to domain wall Joule heating. The scale bar represents 20 µm. (c) Higher resolution scans corresponding to the green and red boxed regions in (b). The polydomain regions contain a high density of closely packed nanodomains. The scale bars represent 2 µm.

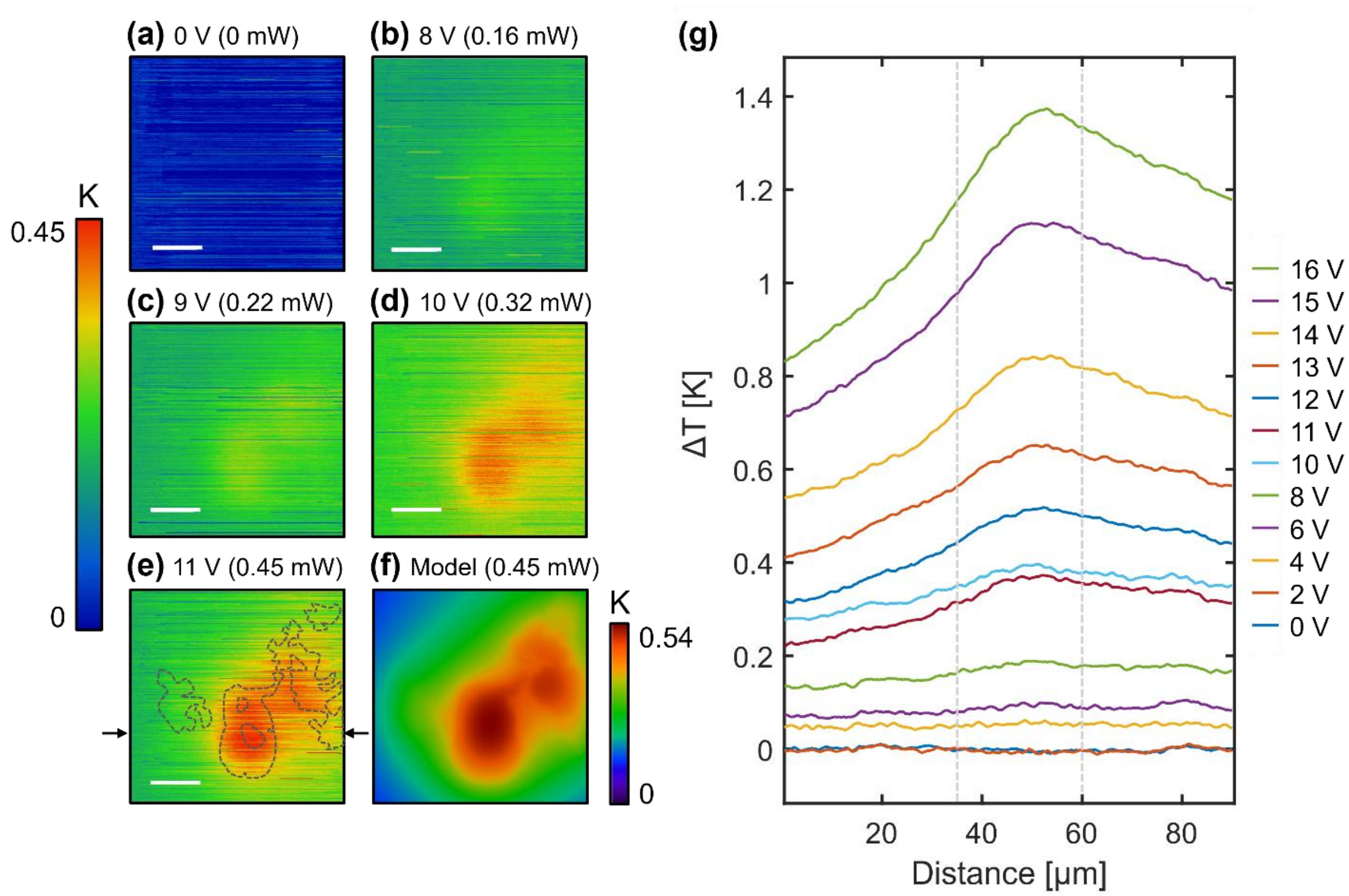

**Figure 4.** (a)-(e) Spatial maps of temperature change obtained as a function of power dissipated, showing the development of hotspots. Outline of domain wall region obtained from piezoresponse force microscopy data is overlaid on (e). Scale bars represent 20 µm. (f) Surface temperature map generated by finite element heat flow modelling. (g) Line sections of temperature for a slow dc voltage sweep taken along the arrowed location indicated in (e). The vertically running dashed lines indicate the edges of the polydomain region. A moving average has also been applied to the data in (g).

Supporting Information

# Direct Thermal Imaging of Domain Wall Hot Spots in $LiNbO_3$

*Lindsey R. Lynch*, J. Marty Gregg, Amit Kumar, Kristina M. Holsgrove and Raymond G. P. McQuaid**

Centre for Quantum Materials and Technologies, School of Mathematics and Physics, Queen's University Belfast, University Road, Belfast BT7 1NN, United Kingdom

E-mail: llynch14@qub.ac.uk, r.mcquaid@qub.ac.uk

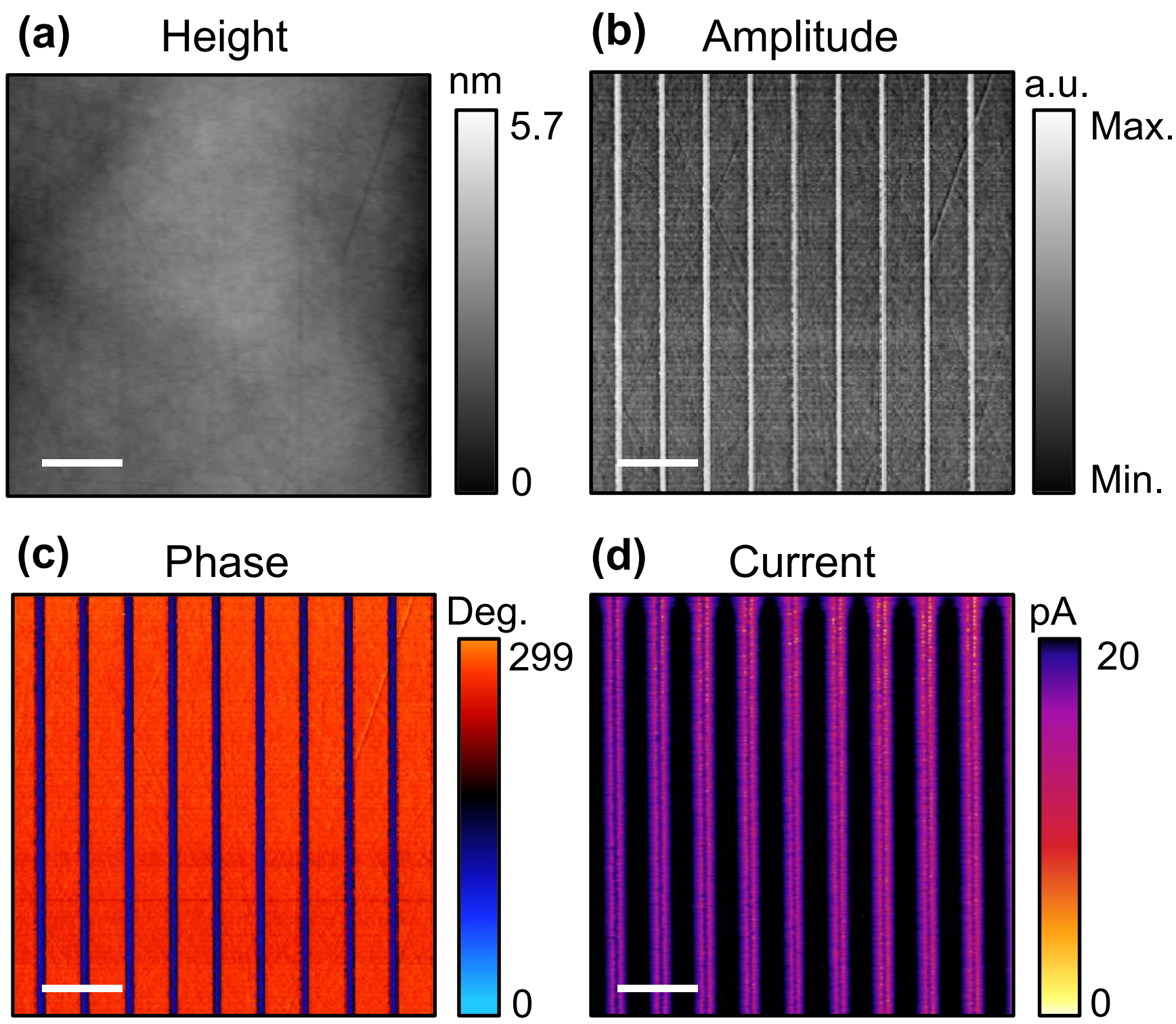

**Figure S1.** (a) Topography map of bare $LiNbO_3$ thin-film surface. (b) Vertical-mode piezoresponse force microscopy amplitude and (c) phase for an engineered domain structure. (d) Current map obtained using the conductive-atomic force microscopy mode, showing that current signals originate from the domain walls. Scale bars represent 3 μm.

**Note 1: Background correction of SThM temperature maps**

In the main manuscript, the temperature data presented in Figure 2b,c and Figure 4a-e have been background corrected to reveal the hot spot morphology. In the raw scanning thermal microscopy (SThM) temperature map, shown in Figure S2a, there are noticeable line-by-line jumps in signal that are highly unlikely to be associated with real temperature changes. These jumps can be seen in the Figure S2b line profile '(1)', where a smooth temperature decay might ordinarily be expected as distance is increased from the buried heat source (the hot spot in the center). Changing tip-sample contact conditions from one scan line to the next (a common issue in scanning probe microscopy methods) is the most likely source of these sporadic signals. However, any changes in tip-sample contact are unknown and challenging to quantify without significant added complexity to the measurement.[1,2] Therefore, our approach is to take the horizontally-running temperature line section along the long edge of the SThM montage as an approximated background correction (indicated by arrow '(2)' in Figure S1a). The idea here is that, if there are any erroneous offsets included in the central line section of temperature (line '(1)' in Fig S2b), then the same erroneous offsets will be captured in the line section taken along the edge of the scan montage (line '(2)' in Fig S2b). This is based on the assumption that the erroneous signal offset remains the same along any given vertically-running SThM line in Figure S2a. Subtracting the temperature line profile '(2)' from '(1)' gives the corrected temperature profile in Figure S2d, which reveals a more smoothly decaying profile away from the location of the hot spot peak temperature (a four-point moving average has been applied also). Further to this, the same background trace can be removed from all line sections of temperature along the bar's long axis, generating the corrected 2D temperature map shown in Figure S2c, where the morphology of the hot spot is more clearly revealed. However, the correction is not perfect: as well as removing erroneous offset signals, any signals associated with real temperature rises along the edge of the bar are also removed. Looking at the background trend in Figure S2b, there is also a slowly varying temperature background rise of ~ 0.2 - 0.4 K, which is removed by the correction alongside the rapidly varying noise signals, likely leading to an underestimation of the hot spot peak temperature in corrected maps. Nonetheless, this approach is a satisfactory compromise for revealing the morphology of the hot spot and the associated temperature rise relative to the immediately surrounding area.

For Figure 4a-e in the main manuscript, the same background correction approach is applied to remove sporadic line-by-line noise. However, in this case the background temperature is quite uniform so an average temperature value is added to each corrected map. This means the SThM maps in Figure 4a-e can be considered as *absolute* temperature rise rather than *relative* temperature rise, as in the case of Figure 2b and c. We also note that four-point moving averaging has been applied to Figure 2c and Figure 4g of the main manuscript but has not been applied to any other line sections in the manuscript.

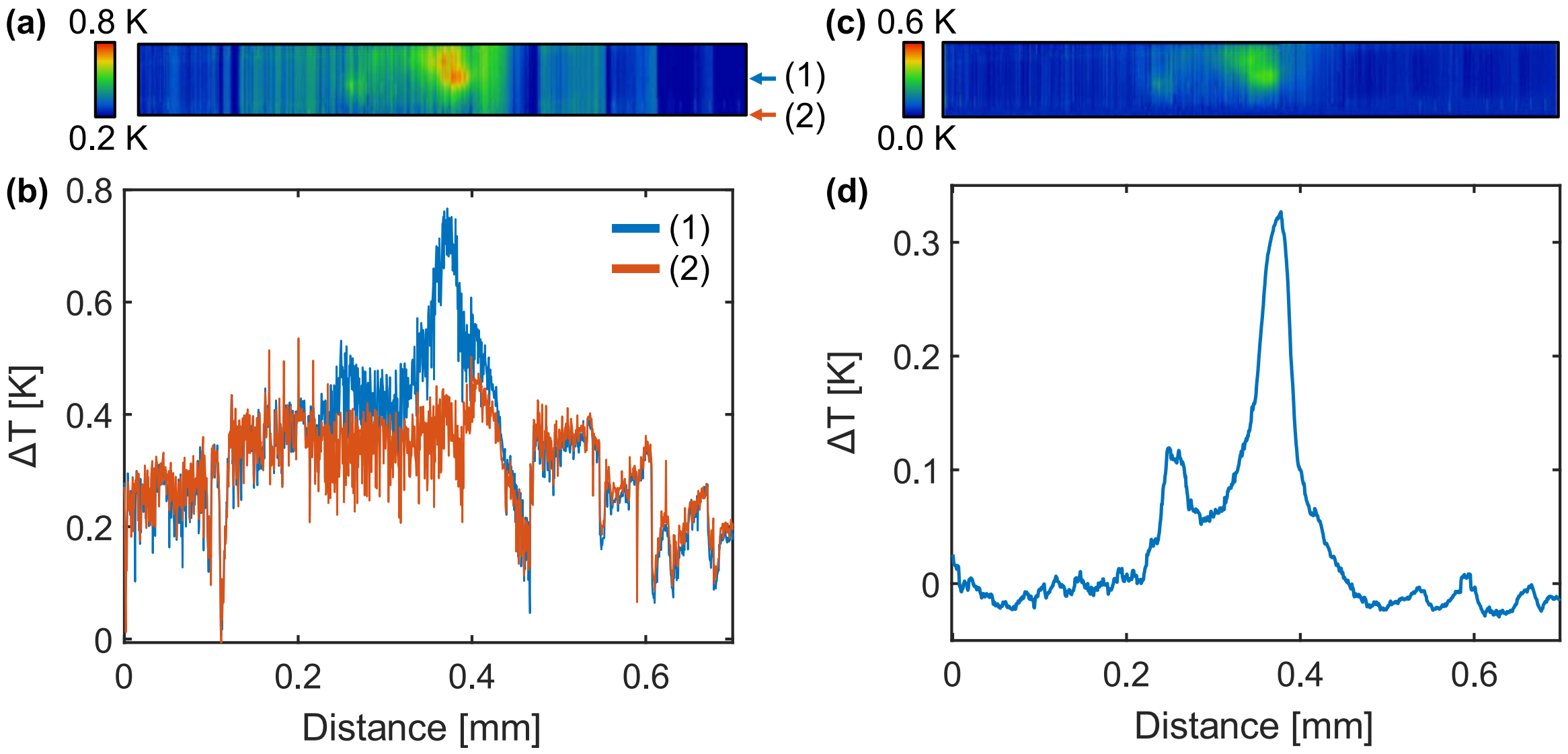

**Figure S2.** (a) As-obtained temperature map on the Pt top electrode measured by scanning thermal microscopy. (b) The blue line '(1)' is a 1D line section of temperature taken along the central axis of the 2D surface temperature map shown in (a), indicated by arrow '(1)'. The orange line '(2)' is taken along the bottom edge of the same 2D temperature map, indicated by arrow '(2)' in (a). The orange line profile is treated as a background signal and removed from each line along the long axis of the 2D map, giving the corrected temperature map in (c). (d) The corrected temperature line section along the central axis which has also been smoothed by a four-point moving average.

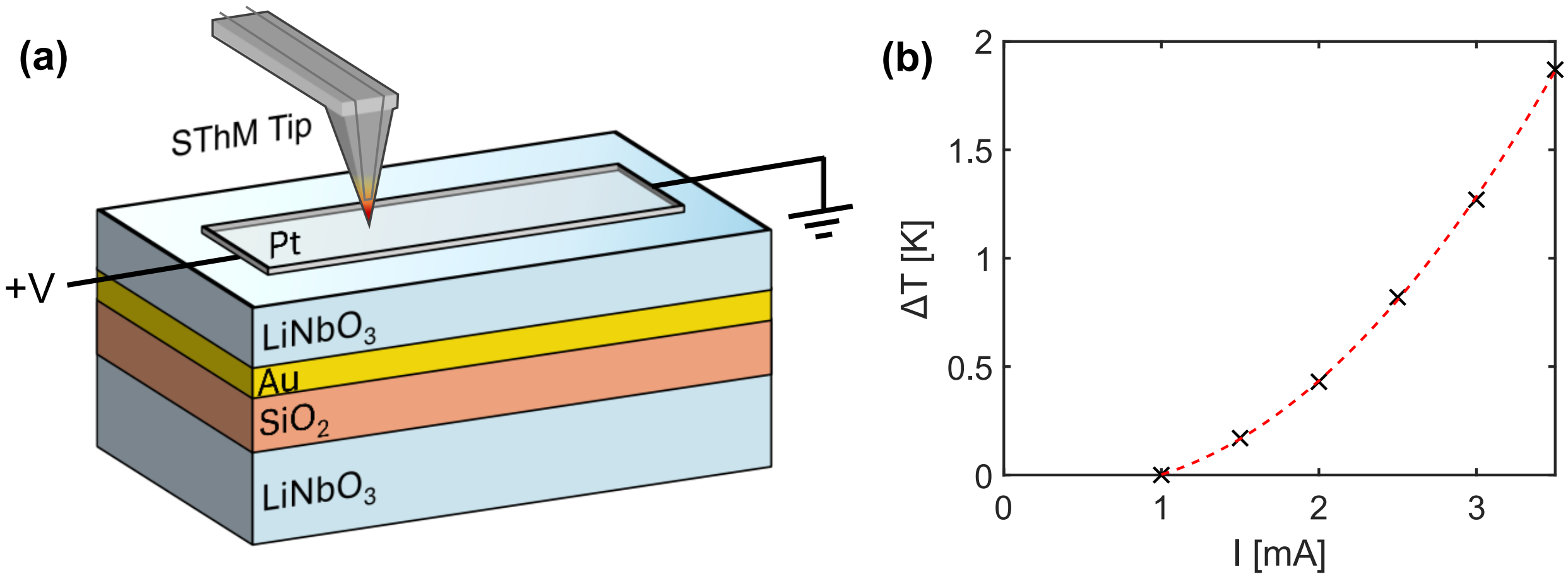

**Figure S3.** (a) Schematic of electrode bar deposited on top of a monodomain $LiNbO_3$ thin-film. Note that the bar is voltage signalled at either end, ensuring current is confined to within the cross section of the bar. (b) The resulting temperature rises measured on the surface of the electrode as a function of current. Note that the maximum current measured in the main manuscript is significantly less than 1 mA and therefore that the contribution from self-heating within the electrode itself is considered to be negligible.

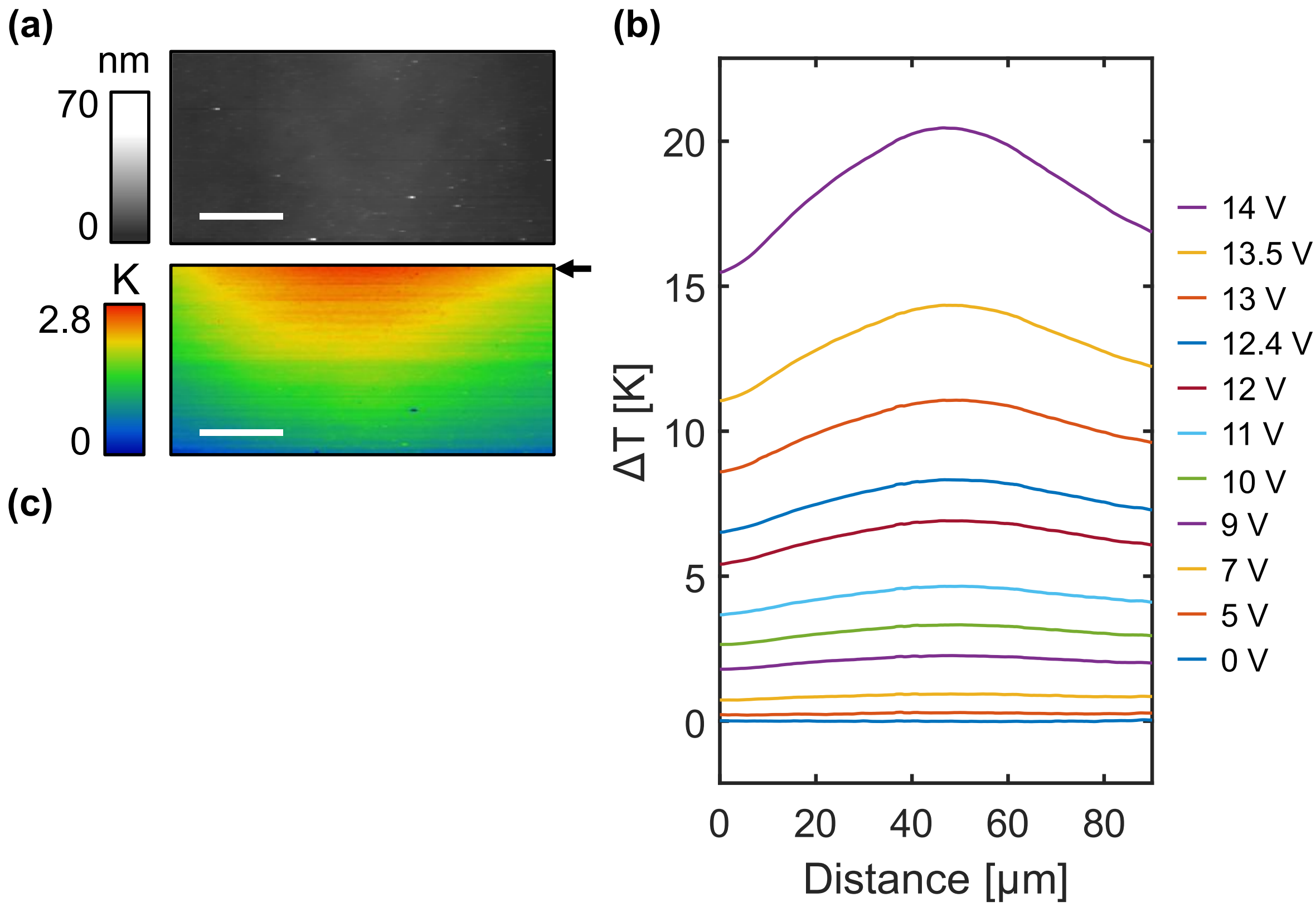

**Figure S4.** (a) Topography (top panel) and temperature map (bottom panel) of a hotspot region on the top electrode in a different domain wall device, under 9 V bias (3 mW power). Scale bars represent 20 µm. (b) Temperature line sections of the hot spot, taken along the arrowed line in (a), examined under increasing bias. (c) Peak hotspot temperature as a function of power dissipated in the device. Temperature rises of 20.5 K are measured on the surface for a current of 2.5 mA, which is driven by a sub-coercive voltage of 14 V. No background corrections have been applied to any data in this figure, all $\Delta T$ values are absolute temperature rises above ambient.

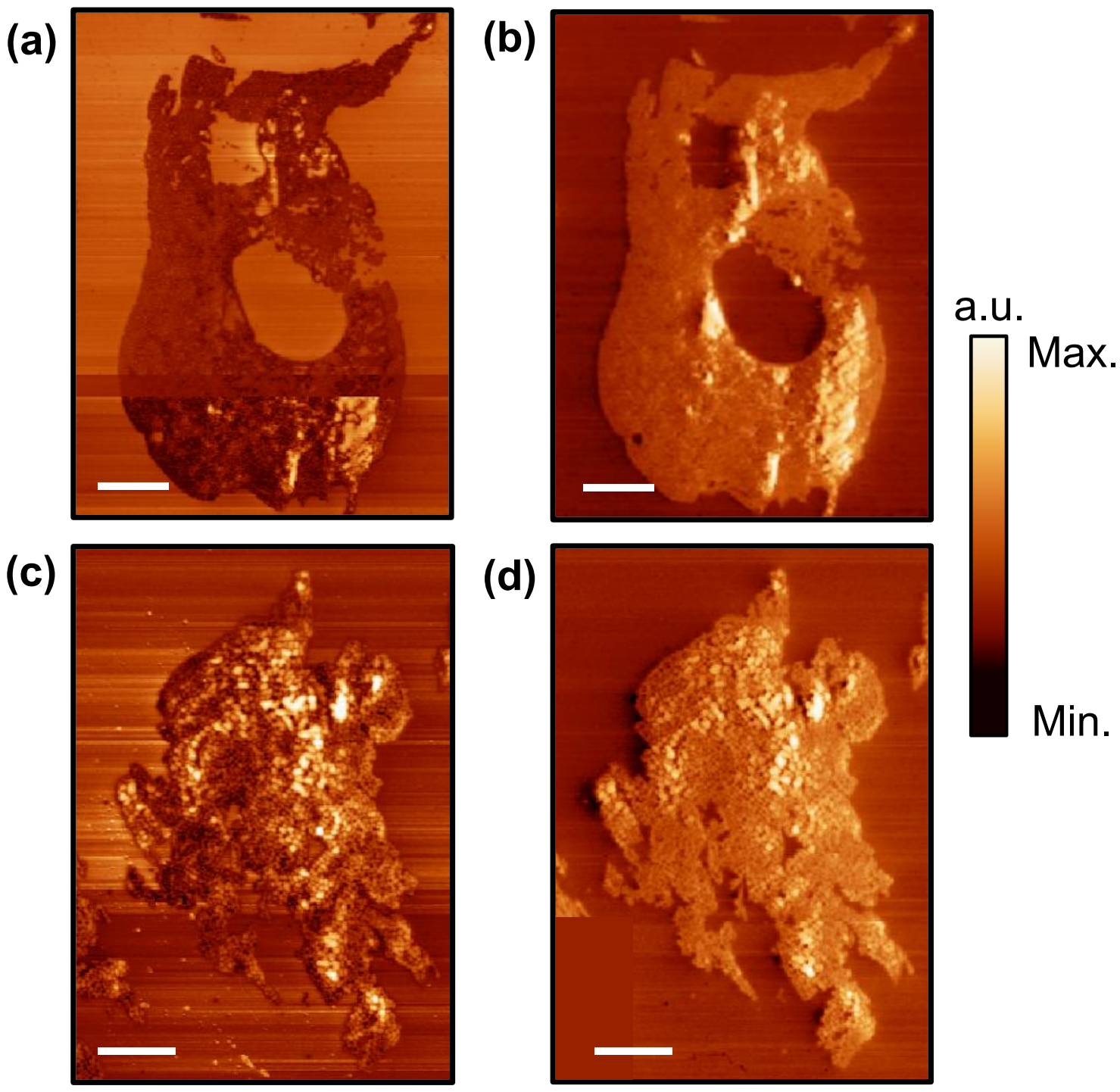

**Figure S5.** Through-electrode piezoresponse force microscopy imaging of the switched domain region responsible for the main hotspot before, (a), and after, (b), extensive thermal investigations. During thermal experiments, the device was subject to varying levels of sub-coercive forward bias on timescales from minutes to hours. Despite changes in the apparent amplitude contrast across domain variants (a common imaging artefact arising from the optical beam deflection measurement of the cantilever)[3], the form of the domain microstructure appears to be unchanged. Scale bars represent 5 μm. Panels (c) and (d) show another region before and after experiment, which similarly remains unchanged. The development of hot spot morphology is therefore primarily driven by increases in dissipated power rather than by changes in the domain microstructure and the associated current leakage pathways. Scale bars represent 3 μm.

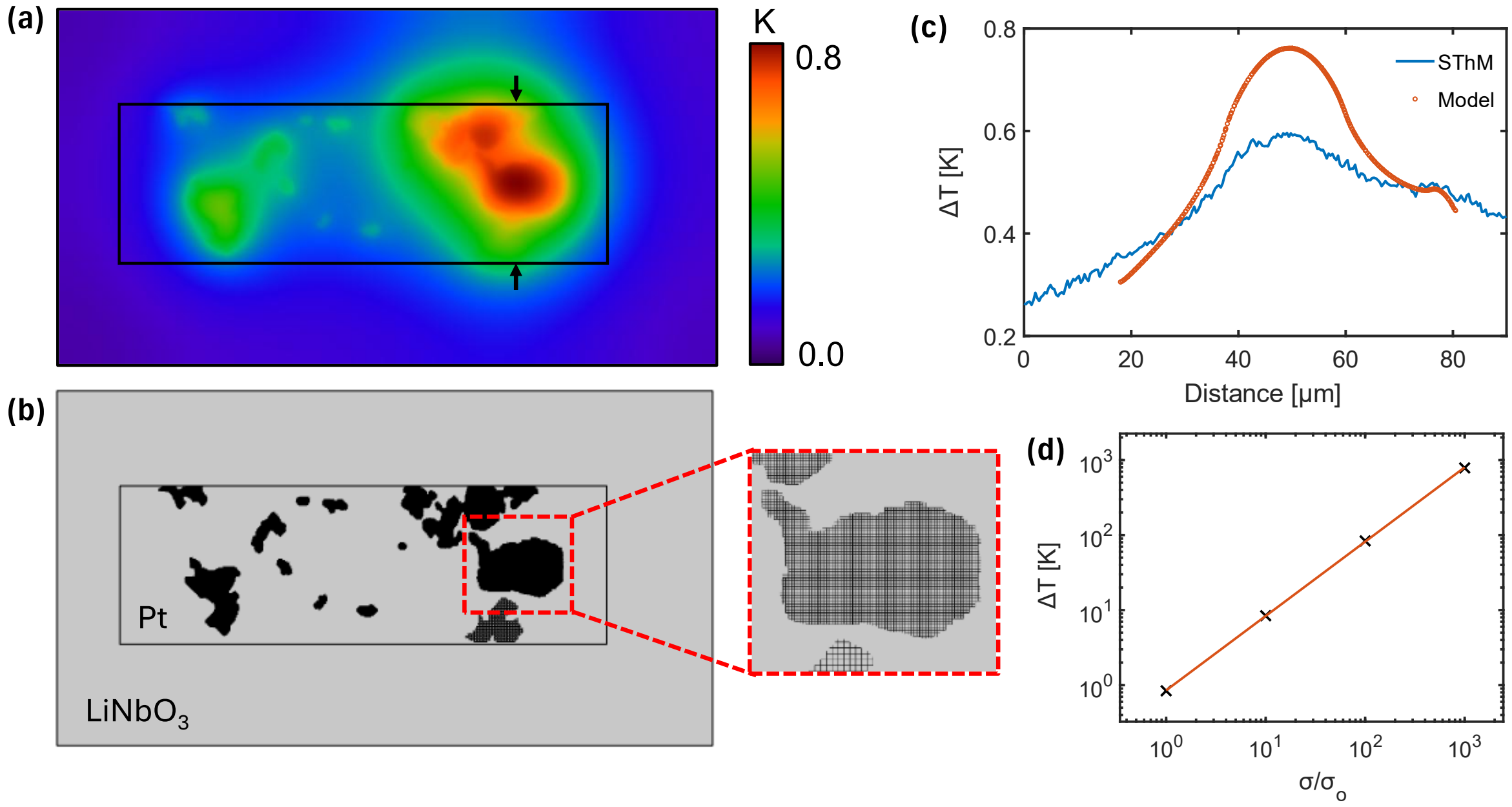

**Figure S6.** (a) Modelled temperature map on the top surface of the sample for the model geometry shown in (b). Inset details square grid arrangement of 2D interfaces (within the $LiNbO_3$ layer) with spacings of 500 nm (central region) and 1 μm (lower region) to approximate the domain wall configuration. (c) Comparison of temperature line section through the hot spot (taken on the surface electrode along the arrowed line section in (a)) with the experimental data, for similar power dissipated. (d) Modelling higher domain wall conductivities results in proportional increases in temperature (due to increased current drawn for equivalent bias).

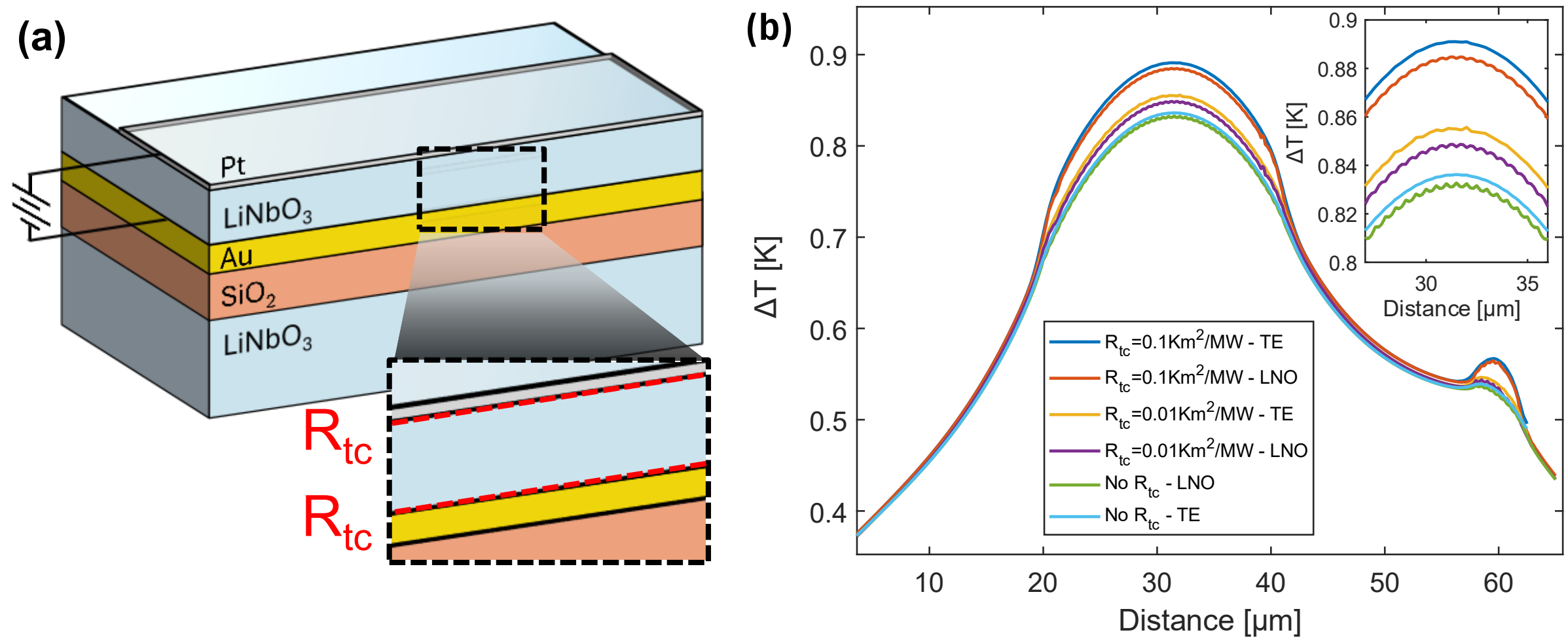

**Figure S7.** (a) Schematic of sample layer geometry with inset indicating where interfacial contact resistances ($R_{tc}$) are included in the model. (b) Line section of modelled temperature taken along the central axis of the wire, intersecting the main hot spot (at distance 30 µm) and a smaller one (at distance 60 µm). The pairs of line sections are taken on the top surface of the electrode (legend TE) and through the center of the $LiNbO_3$ film (legend LNO), so that the discrepancy between temperatures in the respective layers, due to $R_{tc}$, can be examined.

**Note 2: Effect of Thermal Boundary Resistance on Domain Wall Temperature**

In nanoscale devices, the effect of reduced dimensions on heat dissipation can mean that the effective thermal resistance of the device becomes dominated by interfacial contributions, leading to significantly increased local temperatures.[4] This is important for resistive switching nanodevices, where interfacial resistances can cause the temperature of buried current-carrying filaments to be increased by several factors compared to the measured surface temperatures.[5] To assess the possible role of thermal boundary resistances ($R_{tc}$) on domain wall temperatures, we repeated the model from Figure S6 and introduced non-zero $R_{tc}$ values for the top electrode (TE)-$LiNbO_3$ interface and for the $LiNbO_3$-bottom electrode (BE) interface. These values are *a priori* unknown, so we explored a range of values of $R_{tc}$ between 0.01 - 0.001 $Km^2MW^{-1}$, which captures the range of typically encountered values in real thin-film interfaces.[4] The purpose of the simulations is to identify whether the true temperature of the buried domain walls could be significantly larger than the surface temperature measured by SThM. From Figure S7, it can be seen that the influence of the thermal boundary resistances is relatively minor, leading to negligible differences between surface temperature and $LiNbO_3$ layer temperature. In these models, the main effect of the $R_{tc}$ is to increase the absolute temperature rise in the $LiNbO_3$ layer (and concurrent surface temperature) by only

~10%. Therefore, the surface temperatures are likely to be a good representation of the interior temperature in the $LiNbO_3$ layer with the dominant heat loss being predominantly through the surrounding film and the supporting substrate.

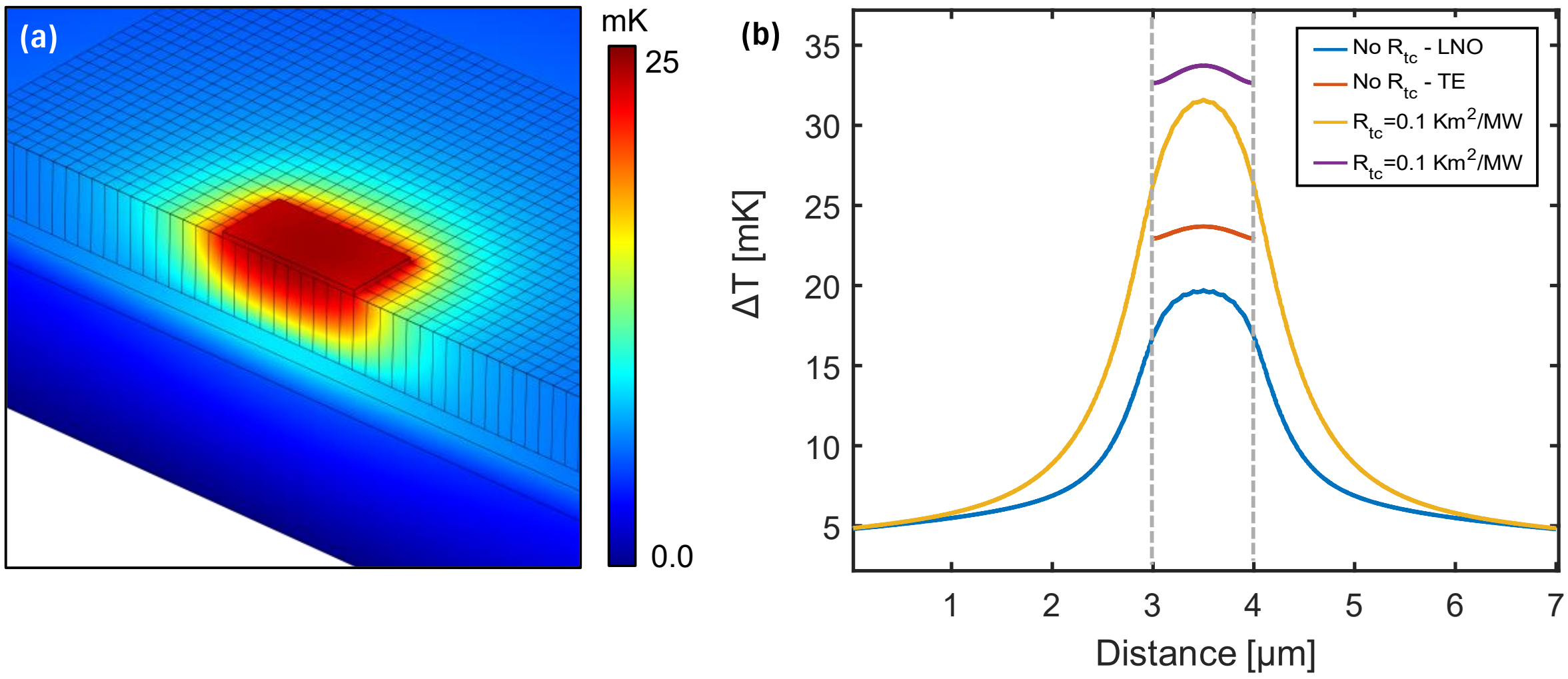

**Figure S8** (a) Modelled temperature map for a 1x1 μm$^2$ lateral dimension device under 5 V bias. The model has been cut cross-sectionally across a plane of symmetry, revealing the interior temperature profile. (b) Line sections of temperature across the top electrode and within the $LiNbO_3$ layer for the cases of no thermal barrier resistances ($R_{tc}$) considered and for the case of $R_{tc}$ = 0.1 Km$^2$/MW applied to the top and bottom $LiNbO_3$/electrode interfaces. The vertically running dashed lines indicate the extent of the electrode on the top surface.

**Note 3: Modelling of 1x1 μm$^2$ $LiNbO_3$ domain wall device**

We have carried out finite element electrothermal modelling of a $LiNbO_3$ device with a much smaller 1x1 μm$^2$ Pt electrode with 40 nm thickness, presented in Figure S8. For computational efficiency, the model geometry was cut along a symmetry axis, and both the $LiNbO_3$ substrate and $SiO_2$ layer thicknesses were reduced by a factor of 10 and their thermal conductivity values increased by the same factor. As before, domain walls were modelled as 2D current carrying planes (using 'electrical shielding' boundary condition), this time arranged in a finer square gridding of 100 nm side (see Figure S8a). To be consistent with the measurements made in the main manuscript, the domain wall electrical conductivity was chosen as $\sigma_0$ = 0.35 S/m. For an applied bias condition of 5 V, the current was 118 nA and the dissipated power was 0.59 μW. The effective current density quoted in the main text is obtained by dividing the current by the surface area of the top electrode, giving a value of 118 nA/μm$^2$. Under these conditions, the peak surface temperature rise is very small, being approximately 20 mK (Figure S8b). We have also carried out the model with thermal boundary resistance value of 0.1 $Km^2/MW$ attributed to both the top and bottom $LiNbO_3$ interfaces, giving a peak temperature rise of 32 mK (i.e. factor ~1.6 increase), which is still negligible.

**Note 4: Calibration of SThM tip temperature**

Calibration measurements were carried out by taking temperature readings on the surface of the platinum bar deposited on the lithium niobate sample. This ensures that material-dependent thermal resistance contributions associated with the tip-sample contact will also be captured in the calibration curve. The sample was silver pasted onto the surface of a heated ceramic surface which was temperature controlled by PID feedback using readout from a resistive platinum sensor. The temperature of the stage was increased incrementally, and enough time was allowed for the tip temperature readout to stabilize before it was recorded. For our Asylum MFP-3D system, the SThM mode is proprietary and the user only has access to the software temperature readout panel (not the Wheatstone bridge output voltage). Plotting the known stage temperature (assumed as the sample surface temperature) versus the tip readout temperature gives the plot shown in Figure S8 for two different thermoresistive SThM probes from the same manufacturer. Tip 1 is the probe used for collecting the data in the main manuscript and the gradient of the plot ($1.8 \pm 0.1$) gives the conversion factor for determining sample surface temperature rise from the software readout value.

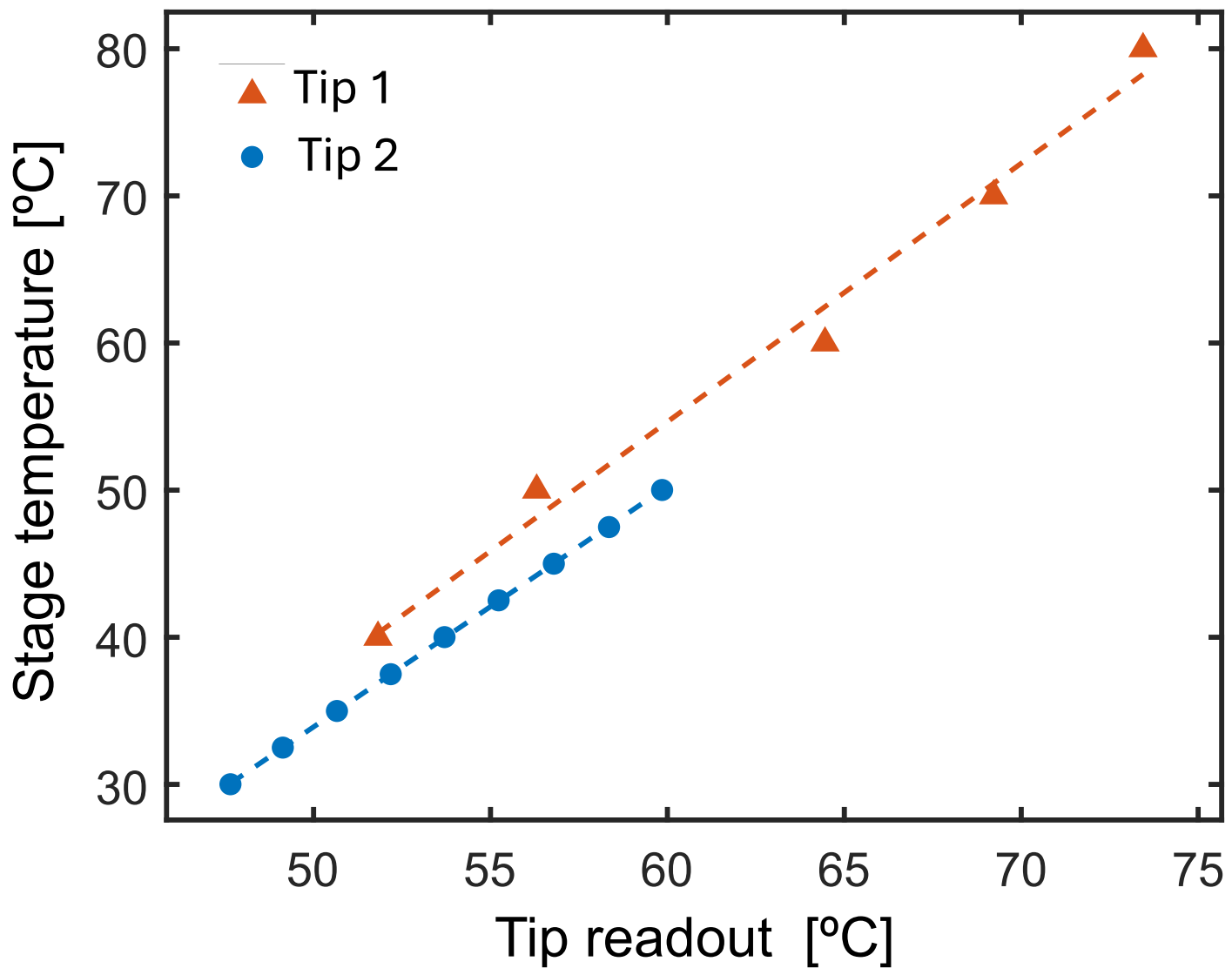

**Figure S9** Plot of temperature of the feedback-controlled stage temperature versus the SThM tip readout for two different commercial KNT-SThM-2an SThM tips.

**Supporting Information References**